\documentclass{article}

\usepackage[a4paper, margin=3cm]{geometry}
\usepackage{authblk}
\usepackage[utf8]{inputenc} 
\usepackage[T1]{fontenc}    
\usepackage{lmodern}
\usepackage{url}            
\usepackage{booktabs}       
\usepackage{amsfonts}       
\usepackage{nicefrac}       
\usepackage{microtype}      
\usepackage{enumitem}       
\usepackage{amsthm}
\usepackage{amsmath}
\usepackage{graphicx}
\usepackage{multirow}
\usepackage[dvipsnames]{xcolor}

\colorlet{col po}{blue!50!black!50}
\colorlet{col rf}{green!70!black}
\colorlet{col mo}{red!90!black}
\colorlet{good red}{red!90!black}
\colorlet{light red}{red!90!black!40}
\colorlet{light gray}{gray!10}
\definecolor{barrier}{gray}{0.45}
\definecolor{darkred}{RGB}{184, 5, 69}
\definecolor{darkgreen}{rgb}{0.0, 0.2, 0.13}
\definecolor{gray}{gray}{0.25}
\definecolor{lightblue}{HTML}{8080FF}
\definecolor{tickgreen}{HTML}{128207}

\definecolor[named]{ACMBlue}{cmyk}{1,0.1,0,0.1}
\definecolor[named]{ACMYellow}{cmyk}{0,0.16,1,0}
\definecolor[named]{ACMOrange}{cmyk}{0,0.42,1,0.01}
\definecolor[named]{ACMRed}{cmyk}{0,0.90,0.86,0}
\definecolor[named]{ACMLightBlue}{cmyk}{0.49,0.01,0,0}
\definecolor[named]{ACMGreen}{cmyk}{0.20,0,1,0.19}
\definecolor[named]{ACMPurple}{cmyk}{0.55,1,0,0.15}
\definecolor[named]{ACMDarkBlue}{cmyk}{1,0.58,0,0.21}

\usepackage{hyperref}
\usepackage{breakurl}

\usepackage{xspace}
\usepackage{pifont}
\usepackage{subcaption}
\usepackage{underscore}
\usepackage{tikz}
\usetikzlibrary{positioning, calc, shapes}

\usepackage{listings}
\definecolor{fencedcolor}{RGB}{52, 52, 255}
\definecolor{codegreen}{rgb}{0,0.6,0}
\definecolor{codegray}{rgb}{0.1,0.1,0.1}
\definecolor{codepurple}{rgb}{0.58,0,0.82}
\definecolor{backcolour}{gray}{0.97}
\definecolor{fencedcolor}{RGB}{52, 52, 255}
\colorlet{fencedBgColor}{CadetBlue!20!White}
\newcommand{\macq}{\ensuremath{\mathbf{acq}}}
\newcommand{\mrel}{\ensuremath{\mathbf{rel}}}
\newcommand{\msc}{\ensuremath{\mathbf{sc}}}
\newcommand{\mrlx}{\ensuremath{\mathbf{rlx}}}

\colorlet{vsyncbarrier}{RubineRed}
\newcommand{\barrierstyle}[1]{\color{vsyncbarrier}\ttfamily\kern-0.1ex\textsubscript{#1}}

\newcommand{\ScBarrier}{\barrierstyle{\msc}}

\lstdefinestyle{mystyle}{
    language=C,
    backgroundcolor=\color{backcolour},   
    commentstyle=\color{codegreen},
    keywordstyle=\color{blue},
    numberstyle=\tiny\color{codegray},
    stringstyle=\color{codepurple},
    basicstyle=\ttfamily\scriptsize,
    breakatwhitespace=false,         
    breaklines=true,                 
    captionpos=b,                    
    keepspaces=true,                 
    numbers=left,                    
    numbersep=5pt,                  
    showspaces=false,                
    showstringspaces=false,
    showtabs=false,                  
    showlines=true,
    tabsize=2,
    morekeywords={inline, template}
}

\newcommand{\sys}{\mbox{\textsc{VSync}}\xspace}

\newcommand{\vsync}{\sys}

\newcommand{\genmcversion}{v0.8}

\newcommand{\eg}{\emph{e.g.}}
\newcommand{\ie}{\emph{i.e.}}

\newcommand{\cmark}{\text{\textcolor{tickgreen}{\ding{51}}}}
\newcommand{\xmark}{\text{\textcolor{good red}{\ding{55}}}}
\mathchardef\hy="2D

\newcommand{\barrier}[1]{\ifmmode\mathit{#1}\else\emph{#1}\fi}

\newcommand{\seqc}{{\color{vsyncbarrier}\msc}}
\newcommand{\rlx}{{\color{vsyncbarrier}\mrlx}}
\newcommand{\acq}{{\color{vsyncbarrier}\macq}}
\newcommand{\rel}{{\color{vsyncbarrier}\mrel}}

\newcommand{\SeqCst}{\seqc\xspace}
\newcommand{\Relaxed}{\rlx\xspace}
\newcommand{\Acquire}{\acq\xspace}
\newcommand{\Release}{\rel\xspace}

\newcommand\dartagnan{\textsc{Dartagnan}\@\xspace}
\newcommand\herd{\textsc{Herd7}\@\xspace}
\newcommand\genmc{\textsc{GenMC}\@\xspace}
\newcommand\arm{\textsc{ARMv8}\@\xspace}
\newcommand\power{\textsc{Power}\@\xspace}
\newcommand\riscv{\textsc{RISC-V}\@\xspace}
\newcommand\xes{\textsc{x86}\@\xspace}

\newcommand\qspinlock{\texttt{qspinlock}\@\xspace}

\title{Verifying and Optimizing Compact NUMA-Aware Locks\\ on Weak Memory Models}
\date{\today}

\author[1]{Antonio~Paolillo}
\author[2]{Hern\'{a}n~Ponce~de~Le\'{o}n}
\author[3]{Thomas~Haas}
\author[1]{Diogo~Behrens}
\author[1]{Rafael~Chehab}
\author[1]{Ming~Fu}
\author[3]{Roland~Meyer}
\affil[1]{{Huawei Dresden Research Center, Germany}}
\affil[2]{{University of the Bundeswehr Munich, Germany}}
\affil[3]{{TU Braunschweig, Germany}}

\date{}

\begin{document}
\maketitle

\begin{abstract}
    Developing concurrent software is challenging, especially if it has to run on modern architectures with Weak Memory Models (WMMs) such as \arm, \power, or \riscv.
    For the sake of performance, WMMs allow hardware and compilers to aggressively reorder memory accesses.
    To guarantee correctness, developers have to carefully place memory barriers in the code to enforce ordering among critical memory operations.

    While WMM architectures are growing in popularity, identifying the necessary and sufficient barriers of complex synchronization primitives is notoriously difficult.
    Unfortunately, publications often consider barriers to be just implementation details and omit them.
    In this technical note, we report our efforts in verifying the correctness of the Compact NUMA-Aware (CNA) lock algorithm on WMMs.
    The CNA lock is of special interest because it has been proposed as a new slowpath for Linux \qspinlock, the main spinlock in Linux.
    
    Besides determining a correct and efficient set of barriers for the original CNA algorithm on WMMs,
    we investigate the correctness of Linux \qspinlock and the latest Linux CNA patch (v15) on the memory models LKMM, \arm, and \power.
    Surprisingly, we have found that Linux \qspinlock and, consequently, Linux CNA are incorrect according to LKMM, but
    are still correct when compiled to \arm or \power. 
\end{abstract}
    
\section{Introduction}
\label{sec:intro}

Correctly developing concurrent software is challenging because of the gigantic number of interleavings of thread executions.
Developing such software to run on modern architectures with Weak Memory Models (WMMs) such as \arm, \power, or \riscv is even more challenging.
WMMs allow hardware and compilers to aggressively reorder memory accesses, which greatly optimizes the performance of applications,
but easily breaks the correctness of concurrent code.

In concurrent code, threads typically communicate via shared variables protected by synchronization primitives (\eg, spinlocks, mutexes, read-write locks).
Such concurrent code should work on WMMs out of the box~\cite{DRF} provided it contains no data races and the synchronization primitives are correct.
Unfortunately, synchronization primitives are not always correct and break in subtle and non-reproducible ways when crucial memory operations are executed out of order.

To enforce some ordering among memory operations, WMMs provide \emph{barriers}, which are either stand-alone explicit fences (\eg, \texttt{DMB ISH} on \arm) or implicit barriers attached to memory operations (\eg, \texttt{LDAR} and \texttt{STLR} on \arm).
Such barriers need to be placed carefully inside the code so that the orderings required by the algorithm are enforced.
However, as concurrent code often lies on the critical path, unnecessary or overly constrained barriers degrade the performance of the whole system.
For this reason, experts spend a great deal of time and effort in identifying the key orderings among memory operations, and optimizing the usage of barriers accordingly~\cite{linux-qspinlock-version4.4, linux-qspinlock-version5.6, linux-qspinlock-version4.16, linux-qspinlock-version4.5, linux-qspinlock-version4.8}.

Identifying the necessary order of memory operations is an error-prone task, even for experts.
For example, the optimization of the barriers in the Linux \qspinlock introduced a bug~\cite{linux-qspinlock-version4.5} that remained unfixed for three years~\cite{linux-qspinlock-version4.16}.
Despite the growing popularity of WMM architectures, new synchronization primitives are often published while omitting the required memory barriers.
For example, Dice and Kogan note the importance of barriers and fences for the implementation of the Compact NUMA-Aware (CNA) lock:
\begin{quote}
``Our actual implementation uses volatile keywords and memory fences as well as padding (to avoid false sharing) where necessarily.''~\cite{cna2019}
\end{quote}
However, they do not provide detailed indications of where these barriers/fences should be placed.
To work correctly on architectures such as \arm, \power, and \riscv, practitioners must find and assign a safe and efficient sequence of barriers to the CNA implementation themselves.

Complementing the work of Dice and Kogan, we report in this technical note our efforts in verifying the correctness of CNA on WMMs and identifying where barriers/fences are necessary.
The identification of barriers is based on the VSync framework~\cite{vsync2021}, which we briefly review in Section~\ref{sec:vsync}.
The verified and optimized CNA lock presented in Section~\ref{sec:cna} has been used in the evaluation of CLoF~\cite{clof2021}.

Since CNA is currently proposed as an alternative implementation of the slowpath of Linux \qspinlock~\cite{lwn2021cna},
we also investigate the correctness of Linux \qspinlock and the latest Linux CNA patch (v15) on the LKMM, \arm, and \power memory models in Section~\ref{sec:linux-cna}.
We found that Linux \qspinlock and, consequently, Linux CNA are incorrect according to LKMM, but both are still correct under the \arm and \power memory models.

\section{VSync Overview}
\label{sec:vsync}

The \vsync framework helps developers to implement, verify, and optimize concurrent code (\eg, CNA lock) on WMMs.
Here, we briefly highlight aspects relevant for this document --- for details on \vsync, please refer to our paper~\cite{vsync2021} and technical report~\cite{vsync-tr}.

\paragraph{Atomics.}
To facilitate the implementation of concurrent code, \vsync provides a library of atomic operations and fences, but it also supports C11 atomics (via \texttt{stdatomic.h}).
The library allows one to implement the atomic primitives with architecture-specific optimizations.
Atomic libraries typically offer two types of barriers: {\em implicit barriers} (\ie, barriers attached to atomic memory operations) and \emph{fences} (\ie, stand-alone barriers, also called {\em explicit barriers}).
Currently, \vsync accepts four barrier modes:
\rlx \space(\emph{relaxed}), \acq \space(\emph{acquire}), \rel \space(\emph{release}), and \seqc \space(\emph{sequentially consistent}). The semantics of barriers can roughly be understood as follows:
\begin{itemize}
	\item no instruction can be reordered \textbf{before} an \textbf{acquire} barrier;
	\item no instruction can be reordered \textbf{after} a \textbf{release} barrier;
	\item a sequentially-consistent read implies an acquire barrier;
	\item a sequentially-consistent write implies a release barrier;
	\item a sequentially-consistent fences implies both an acquire barrier and a release barrier;
	\item sequentially-consistent accesses cannot be reordered with each other.
\end{itemize}
The relaxed mode \Relaxed allows the processor to perform all memory-access optimizations, whereas the sequentially consistent mode \SeqCst disables most optimizations.
The modes \Release and \Acquire are between the other two, allowing some optimizations.
They are used to correctly implement efficient message-passing, producer/consumer patterns, and, of course,
lock and unlock internals.

\paragraph{Verification.}
There are two key properties that synchronization primitives such as CNA lock should satisfy:
{\em mutual exclusion}, \ie, no two threads can execute the critical section (protecting the shared variables) at the same time;
and {\em await termination}, \ie, threads eventually leave the \lstinline|lock_acquire()| and \lstinline|lock_release()| functions.
To verify these properties on WMMs, \vsync employs a model checker --- more specifically, AMC~\cite{vsync2021} integrated in GenMC~\cite{genmc-gh,GenMC} \genmcversion{} or later.
This model checker is aware of the \vsync library and can therefore correctly reason about library calls.

\vsync compiles to LLVM-IR the synchronization primitive together with a special client code that creates threads and drives the execution.
The client code used with CNA lock (see Figure~\ref{fig:client-code}) increments a variable \texttt{v} in the critical section.
Mutual exclusion is guaranteed if the model checker can show that \texttt{v} equals $N$ at the end of all possible executions of the client code.
The implementation is \emph{live} (guarantees await termination) if the model checker can show that all possible executions of the client code terminate.

\begin{figure}
    \centering
\begin{minipage}[T]{0.42\linewidth}
    \centering
    \input{figures/client-code}
    \caption{CNA client code used for checking and barrier optimization.}\label{fig:client-code}
\end{minipage}\hspace{1cm}
\begin{minipage}[T]{0.42\linewidth}
    \centering
        \input{figures/cna-opt-report}
    \caption{%
        \vsync optimization report with maximally-relaxed barrier combination.
    }\label{fig:vsyncer-report}
\end{minipage}
\end{figure}

\paragraph{Optimization.}
{\em Barrier optimization} is the task of finding a combination of maximally-relaxed barrier modes for the concurrent code at hand, that still guarantees the two above key properties.
This maximally-relaxed barrier combination allows the hardware to perform all memory-access optimizations that do not introduce property-violating behavior.
Note that the combination cannot, however, enforce that the hardware indeed does optimizations and that the system performs better or is faster.

\vsync optimizes barriers by gradually relaxing their modes, while employing the model checker at each step to verify the correctness of the current barrier combination.
The order in which barriers are relaxed is decided by the adaptive linear relaxation algorithm, which allows \vsync to quickly find a maximally-relaxed barrier combination and produce a report as in Figure~\ref{fig:vsyncer-report}.
The reported barrier combination is guaranteed to be both correct and \emph{maximally relaxed}, \ie, further relaxing any of the barriers would cause the implementation to fail on WMMs.

\paragraph{Memory model.}
The default memory model is the intermediate memory model~\cite{podkopaev2019bridging} (IMM), which unifies most dependency-tracking WMMs, including \xes, \power, \arm, and \riscv.
We are currently working to support the Linux Kernel Memory Model~\cite{lkmm-openstd} (LKMM) in \vsync.

\paragraph{Caveat.}
Model checking can only analyze the correctness relative to the given client code.
Our client code (Figure~\ref{fig:client-code}) is parametric in the number $N$ of threads.
It is up to the user to find a sufficiently large parameter $N$ for which the client code can fully exercise the code under consideration and,
in turn, produce a dependable verification and optimization result.
We verified the CNA lock using 3, 4, and 5 threads.
\vsync finds the same maximally-relaxed barrier combination with 4 and 5 threads.

\section{Verification and Optimization of CNA lock on WMMs}
\label{sec:cna}

In this section, we report the results of our verification and optimization of the original CNA algorithm~\cite{cna2019} on WMMs.
We also discuss one example of how relaxing the resulting barrier combination can cause a bug,
showing that the resulting combination is indeed \emph{maximally-relaxed}.

\begin{figure}[t]
    \hspace{-8mm}
    \begin{subfigure}[t]{79mm}
\begin{lstlisting}[style=mystyle]
void cna_lock(cna_lock_t *lock, cna_node_t *me) {
  me->next = 0;<@\label{cnacode:initnode1}@>
  me->socket = -1;
  me->spin = 0;<@\label{cnacode:initnode2}@>

  cna_node_t * tail = SWAP<@\ScBarrier@>(&lock->tail, me);<@\label{cnacode:swap}@>
  if (! tail ) { me->spin = 1; return; }<@\label{cnacode:lockfastexit}@>

  me->socket = current_numa_node();
  tail->next =<@\RelBarrier@> me;

  while (!me->spin<@\AcqBarrier@>) { CPU_PAUSE(); }<@\label{cnacode:lockspin}@>
}

void cna_unlock(cna_lock_t *lock, cna_node_t *me) {
  if (!me->next<@\AcqBarrier@>) {<@\label{cnacode:unlockif1}@>
    if (me->spin == 1) {
      if (CAS<@\ScBarrier@>(&lock->tail, me, NULL) == me) return;
    } else {
      cna_node_t *secHead = (cna_node_t *) me->spin;
      if (CAS<@\ScBarrier@>(&lock->tail, me, secHead->secTail) == me) {
        secHead->spin =<@\RelBarrier@> 1;
        return;
      }
    }
    while (me->next == NULL) { CPU_PAUSE(); }
  }
  cna_node_t *succ = NULL;
  if (keep_lock_local() && (succ = find_successor(me))) {<@\label{cnacode:unlockif2}@>
    succ->spin =<@\RelBarrier@> me->spin;<@\label{cnacode:rlxbar1}@>
  } else if (me->spin > 1) {
    succ = (cna_node_t *) me->spin;
    succ->secTail->next = me->next;
    succ->spin =<@\RelBarrier@> 1;
  } else {
    me->next->spin =<@\RelBarrier@> 1;
  }
}
\end{lstlisting}
    \end{subfigure}\hspace{2em}
    \begin{subfigure}[t]{79mm}
    \begin{lstlisting}[style=mystyle, firstnumber=45]
cna_node_t * find_successor(cna_node_t *me) {
  cna_node_t *next = me->next;
  int mySocket = me->socket;

  if (mySocket == -1) mySocket = current_numa_node();
  if (next->socket == mySocket) return next;

  cna_node_t *secHead = next;
  cna_node_t *secTail = next;
  cna_node_t *cur = next->next<@\AcqBarrier@>;

  while (cur) {
    if (cur->socket == mySocket) {
      if (me->spin > 1)
        ((cna_node_t *)(me->spin))->secTail->next = secHead;
      else
        me->spin = (uintptr_t )secHead;
      secTail->next = NULL;
      ((cna_node_t *)(me->spin))->secTail = secTail;
      return cur;
    }
    secTail = cur;
    cur = cur->next<@\AcqBarrier@>;
  }
  return NULL;
}
\end{lstlisting}
\end{subfigure}
    \caption{
        Code implementing CNA Lock as presented in the original work~\cite{cna2019},
        enriched with barrier annotations.
        The annotations {\Acquire}, {\Release} and {\SeqCst} respectively mean
        an acquire, release and sequentially-consistent implicit barrier.
        If the annotation is (1) after a \texttt{=} sign, it concerns a write access;
        (2) after an expression, it concerns a read access;
        or (3) after a function name, it concerns an atomic primitive.
        The absence of a barrier annotation means, for a struct field, a relaxed access, and for a non-shared stack variable, a non-atomic access.
    }
    \label{fig:cnacode}
\end{figure}

Using the code provided in the original CNA work~\cite{cna2019},
we identified all concurrent accesses and replaced them by the corresponding atomic operations with \SeqCst barrier modes.
That concerns not only the \texttt{SWAP} and \texttt{CAS} operations, but also every read and write access to the fields of \texttt{cna\_node\_t} and \texttt{cna\_lock\_t}.
Figure~\ref{fig:cnacode} depicts the implementation with the resulting barrier relaxations, verified and optimized by \vsync.
For the sake of readability, we use a shortened notation to identify the necessary barrier modes: 
operations marked with subscripts
\Acquire,
\Release, and
\SeqCst have the respective barrier mode.
Other operations have no barrier attached (\ie, \Relaxed).
These barriers guarantee mutual exclusion and await termination of the CNA locking algorithm on WMMs.

As mentioned above, we ran our client code configured with 3, 4, and 5 threads.
We report the different checking and optimization times in Table~\ref{tab:totaltimes}.

\begin{table}
\caption{Checker and optimizer time for CNA.}
\label{tab:totaltimes}
\centering
\begin{tabular}{rrr}
\toprule
\textbf{Threads}     & \textbf{Verification time} & \textbf{Optimization time} \\
\multicolumn{1}{l}{} & \multicolumn{1}{l}{}       & \multicolumn{1}{l}{}       \\
\midrule
3                    & 1s                         & 21s                        \\
4                    & 1m31s                      & 1m58s                      \\
5                    & 1h49m54s                   & 2h00m11s                   \\
\bottomrule
\end{tabular}
\end{table}

\paragraph{Example of a Bug Caused by Overly-Relaxed Barriers.}

\vsync guarantees the provided solution is \emph{``maximally-relaxed''}, meaning
any further relaxation could lead to a property-violating execution.
To illustrate this point, we try to relax a barrier from the code depicted in Figure~\ref{fig:cnacode},
and explain how it leads to data corruption.
On Line~\ref{cnacode:rlxbar1}, we remove the release barrier, fully relaxing the write to \texttt{succ->spin}.
With that simple change, our \vsync checker detects an \emph{unsafe} execution with as few as two threads:
mutual exclusion of the critical section is violated, allowing both threads to modify data concurrently.

To see the issue, let us use the following scenario where two threads $T_0$ and $T_1$, with a queue initially empty (\texttt{lock->tail == NULL}), run their code as depicted in Figure~\ref{fig:client-code}: both threads call \texttt{cna\_lock}, increment a shared counter \texttt{v} (which is not of an atomic type), and call \texttt{cna\_unlock}.

The critical section consists of an increment of the variable \texttt{v}.
In the C language, this requires a \emph{read} of the current value of \texttt{v} from memory, followed by a write of value \texttt{v+1} back to memory.
If mutual exclusion of the critical section is violated, both threads could potentially read the same value \texttt{v} (in this case \texttt{v=0}), and both would then write \texttt{1} to the variable \texttt{v}.

Assume $T_0$ executes \texttt{cna\_lock} first: it initializes \texttt{node[0]} (Lines~\ref{cnacode:initnode1}-\ref{cnacode:initnode2}), adds itself as the tail to the list using the \texttt{SWAP\ScBarrier} operation (Line~\ref{cnacode:swap}), and, as the list was empty, sets its own \texttt{node[0]->spin} to 1 and returns (Line~\ref{cnacode:lockfastexit}), entering the critical section.

In parallel, $T_1$ executes \texttt{cna\_lock} as well and adds itself as the tail after
$T_0$; as $T_0$ is already in the queue, \texttt{node[0].next} is now \texttt{node[1]}, and $T_1$ starts to spin on its own \texttt{node[1]->spin} field (Line~\ref{cnacode:lockspin}), waiting for $T_0$ to release it.

However, on $T_0$'s side, the read of \texttt{v} in the critical section occurs (\texttt{v} evaluates to 0) but the write operation is delayed because
there is no release barrier to avoid reordering it with later instructions.
Indeed, the first test in \texttt{cna\_unlock} fails (Line~\ref{cnacode:unlockif1}) as \texttt{node[0]->next} is not null but is equal to \texttt{\&node[1]}, therefore
the body of the \texttt{if} is skipped, and no barrier contained in it has any impact on that execution.
After that, the execution jumps to Line~\ref{cnacode:unlockif2}:
the function \texttt{keep\_lock\_local()} returns true (as it does most of the time),
so \texttt{succ} is initialized to the return value of \texttt{find\_successor(\&node[0])}, which is \texttt{\&node[1]}.
No barrier in the \texttt{find\_successor} function prevents further delaying the write of the critical section.
Therefore, we enter the body of the \texttt{if}.
Previously, Line~\ref{cnacode:rlxbar1} included a release barrier, thus forbidding to further delay the critical section write (of \texttt{v+1}).
However, we relaxed it in this example, allowing the atomic write to \texttt{succ->spin} (Line~\ref{cnacode:rlxbar1}) to occur first, thus releasing $T_1$ from spinning (Line~\ref{cnacode:lockspin}).
$T_1$ will then read \texttt{v} before $T_0$ had the chance to write \texttt{1} to it (thus violating mutual exclusion), so they both read \texttt{0}.
Consequently, both threads will write \texttt{1} to \texttt{v}, resulting in data corruption for this specific execution.
Our model checker (GenMC with AMC) is able to detect this execution as it violates an assertion that ensures the value of \texttt{v} to be correct (see client code in Fig.~\ref{fig:client-code}).

With this example, we showed that in an implementation with wrong barriers,
subtle reorderings can occur and, combined with concurrent execution,
can lead to data corruption.
The safety property is thus violated.

We ran both this example and the correct version on an \arm computer:
in some cases, we observed data corruption in the former whereas we never observed any hang nor failure with the version that has correct and maximally-relaxed barriers (the one depicted in Figure~\ref{fig:cnacode}).

\section{Verification of Linux CNA on LKMM}
\label{sec:linux-cna}

The CNA lock is about to be merged into the Linux kernel as an alternative implementation for the slowpath of \qspinlock~\cite{lwn2021cna}.
The verification of the CNA lock in user space (Section~\ref{sec:cna}) only partially supports the correctness of the CNA version being introduced in Linux.
There are two reasons for that.
First, the CNA algorithm has evolved to match Linux requirements.
In particular, it is combined with a \qspinlock, which in turn has its own memory barriers that can affect CNA.  
Second, the memory model used in \sys is IMM, but Linux has its own memory model, LKMM.
As mentioned in Section~\ref{sec:vsync}, \sys does not yet fully support LKMM, so we decided
to use the \dartagnan~\cite{dartagnan-gh, dartagnan} tool to verify (but not optimize) CNA under LKMM.
\dartagnan is a Bounded Model Checking (BMC) tool that can find concurrency bugs in programs executed under various memory models, and in particular also under LKMM.
However, it comes with minor limitations which we discuss below.\footnote{%
Note that a previous version of this technical note~\cite{DBLP:journals/corr/abs-2111-15240} reported in this section results using the \genmc tool.
Unfortunately, those results were incorrect due to code simplifications and model checker issues with LKMM.
The results presented here supersede the previous results.
}

\begin{figure}[!htp]
    \hspace{-8mm}
    \begin{subfigure}[t]{79mm}
\begin{lstlisting}[style=mystyle]
typedef struct qspinlock {
  int spinlock;     /* cmpxchg-based spin lock */
  union {
    atomic_t val;
        /* val is only CNA tail
         * not a union with locked and pending */
    };
};

void
queued_spin_lock_slowpath(struct qspinlock *lock, u32 val)
{
  struct mcs_spinlock *prev, *next, *node;
  u32 old, tail;
  int idx;
  /* completely removed lock-pending logic */

queue:
  node = this_cpu_ptr(&qnodes[0].mcs);
  idx = node->count++;
  tail = encode_tail(smp_processor_id(), idx);
  node = grab_mcs_node(node, idx);

  barrier();

  node->locked = 0; <@\label{qspinlockcode:init}@>
  node->next = NULL;
  cna_init_node(node);

  smp_wmb(); <@\label{qspinlockcode:wmb}@>

  old = xchg_tail(lock, tail);
  next = NULL;

  if (old & _Q_TAIL_MASK) {
    prev = decode_tail(old);
    WRITE_ONCE(prev->next, node);

    arch_mcs_spin_wait(&node->locked);

    next = READ_ONCE(node->next);
    if (next)
      prefetchw(next);
  }
  if ((val = cna_wait_head_or_lock(lock, node)))
    goto locked;

  /* removed lock pending logic in the next read */
  val = atomic_read_acquire(&lock->val);
  
locked:
  /* acquire spinlock after acquiring CNA lock */ 
  await_while(cmpxchg_acquire(&lock->spinlock, 0, 1) != 0);

  if (cna_try_clear_tail(lock, val, node))
    goto release;

  if (!next)
    next = smp_cond_load_relaxed(&node->next, (VAL));

  cna_lock_handoff(node, next); <@\label{qspinlockcode:handoff}@>

release:
  __this_cpu_dec(qnodes[0].mcs.count);
}
bool __try_clear_tail(struct qspinlock *lock,
               u32 val,
               struct mcs_spinlock *node)
{
  /* removed pending bits logic */
  return atomic_try_cmpxchg_relaxed(&lock->val, &val, 0);
}
void queued_spin_unlock(struct qspinlock *lock)
{
  /* release spinlock */ 
  smp_store_release(&lock->spinlock, 0);
}
    \end{lstlisting}
    \caption{Modified fast path in {\tt qspinlock.\{c,h\}}}\label{fig:linux-cnacode-qspin}
  \end{subfigure}\hspace{2em}
\begin{minipage}[t]{79mm}
  \begin{subfigure}[t]{\linewidth}
    \begin{lstlisting}[style=mystyle, firstnumber=45]
bool probably(unsigned int num_bits)
{
  return 0;
}

bool intra_node_threshold_reached(struct cna_node *cn)
{
  return my_threshold != 0;
}

u32 cna_wait_head_or_lock(struct qspinlock *lock,
             struct mcs_spinlock *node)
{
  // ... (truncated)
  if (!cn->start_time || !intra_node_threshold_reached(cn)) {
    if (cn->numa_node == CNA_PRIORITY_NODE)
      cn->numa_node = cn->real_numa_node;

    /* forced only two cna_order_queue call by replacing loop:
     *  while (LOCK_IS_BUSY(lock) && !cna_order_queue(node))
     *	  cpu_relax();
     * with these calls: */
    cna_order_queue(node);
    cna_order_queue(node);

  } else {
    cn->start_time = FLUSH_SECONDARY_QUEUE;
  }
  return 0;
}
    \end{lstlisting}
    \caption{Minor changes in {\tt qspinlock_cna.h} }\label{fig:linux-cnacode-changes}
  \end{subfigure}
  \vspace{10mm}

  \begin{subfigure}[t]{\linewidth}
    \begin{lstlisting}[style=mystyle]
#define N 4 // number of threads
struct qspinlock lock;
struct cna_node my_node[N];
int my_threshold = 0;
int x = 0; // first counter
int y = 0; // second counter
void run(int id) {
  queued_spin_lock_slowpath(&lock, 0);
  // critical section
  x = x + 1;
  y = y + 1;
  queued_spin_unlock(&lock); 
}
void main() {
  pthread_t t[N];
  for (int i=0; i < N; i++)
    pthread_create(&t[i], 0, run, i);
  my_threshold = 1; // non-deterministically decide when
                    // threshold is reached
  for (int i=0; i < N; i++)
    pthread_join(t[i], 0);
  assert (x == y);  // check of mutual exclusion
}
\end{lstlisting}
    \caption{Client code for Linux CNA}\label{fig:linux-cnacode-client}
  \end{subfigure}
\end{minipage}
  \caption{
    Linux CNA and \qspinlock modifications and client code to verify Linux CNA with \dartagnan on LKMM, \arm and \power.
  }\label{fig:linux-cnacode}
\end{figure}

Since the CNA lock is combined with \qspinlock, we first verified the correctness of the latter in isolation.
During this process, \dartagnan found several correctness violations.
According to LKMM, \qspinlock guarantees neither mutual exclusion nor await termination, and it contains a data race.
Below we discuss all these issues.

This might initially be shocking: \emph{how is it possible that nobody noticed the main locking algorithm of the kernel to be fundamentally broken?}
The reason for this is that the lock is incorrect according to LKMM, but all these issues disappear when the code is compiled and run on hardware.
This discrepancy is due to the fact that language-level memory models (like LKMM) allow more flexible reorderings
than hardware-level memory models (like \arm and \power), which are effectively implemented in hardware.

Besides supporting code verification under LKMM, \dartagnan also allows the user to compile\footnote{
    \dartagnan follows the kernel implementation of atomics,
    \ie, located in the kernel source tree in \colorbox{black!10}{\texttt{arch/<target-arch>/include/asm/atomic.h}}.
}
the code to different architectures, and verify the resulting assembly output under the corresponding hardware memory model.
Using \dartagnan we were able to prove the correctness\footnote{
    As \dartagnan is a BMC tool, loops are unrolled.
	The correctness holds up to the unrolling bound.}
(both in terms of safety and await termination) of \qspinlock under \arm and \power.

The CNA lock inherits the correctness issues we found in \qspinlock, making it incorrect as well under LKMM.
However, using two changes described below, we were able to verify both \qspinlock and CNA under LKMM.
We next tried to verify the full CNA implementation under \arm and \power.
The verification of Linux CNA (patch v15~\cite{cnalinuxv15}) with 4 threads took less than 16 hours under LKMM and less than 12 hours under each targeted architecture (see Table~\ref{tab:veriftime} for detailed times for each verification case).
We detailed on Github the procedure to reproduce the different runs of the model checker,
together with a Linux kernel patch required to make it work~\cite{cnaverification-gh}.
As we describe next, we could only perform a partial verification of the Linux CNA due to some limitations in \dartagnan.

\paragraph{\dartagnan limitations.}
We faced a few challenges while verifying Linux CNA.
First, \dartagnan does not support mixed-sized accesses.
Linux \qspinlock combines the fast spinlock with the tail of the MCS lock or CNA lock in a single 32-bit variable.
This variable is accessed with atomic operations of multiple widths: times 32-bit, times 8-bits.
\dartagnan cannot work with that.
Second, the execution time of \dartagnan is exponential in the number of threads and the size of the code.
Since CNA lock is used in the slow path only, more threads are required to exercise its code than when CNA is used as originally proposed (Section~\ref{sec:cna}).
The higher number of threads and the additional complexity introduced by other CNA changes drastically increase verification time.

\paragraph{CNA and \qspinlock changes.}
To mitigate the limitations mentioned above, we removed the pending bit strategy of \qspinlock and split the fast path spinlock and the CNA lock tail in two distinct 32-bit fields.
We also changed the \qspinlock algorithm as follows (see Figure~\ref{fig:linux-cnacode-qspin}).
A thread first acquires the slow path lock, \ie, the CNA lock, as if the fast path spinlock acquisition would have failed.
Once the thread has acquired the CNA lock, it reorders or flushes the CNA queue, as in the original Linux CNA patch.
After that, the thread acquires the fast path spinlock (without pending bit logic).
The remainder follows the original Linux CNA patch.
The thread releases the CNA lock and enters the critical section while holding the fast path spinlock.
After the critical section, the thread simply releases the fast path spinlock.
In the CNA algorithm itself, the required changes were minimal (see Figure~\ref{fig:linux-cnacode-changes}).
To simplify verification, we disable the shuffle reduction by returning always 0 from the \texttt{probably()} function.
Moreover, we call \texttt{cna\_order\_queue()} only twice.
Finally, the decision of when the intra-node threshold is reached is performed non-deterministically in the client code (Figure~\ref{fig:linux-cnacode-client}).
We expect that these changes do not affect the lock correctness whereas they reduce
the complexity of the implementation such that verification times become reasonable.

\paragraph{Correctness violations in \qspinlock.}

\dartagnan found three issues with \qspinlock under LKMM:
it does not guarantee mutual exclusion (safety violation), it contains a data race (also confirmed by \genmc), and it can hang (await termination violation).
To showcase these issues, we present litmus tests that isolate the problematic behaviors and hence can be analyzed, and thus confirmed, by the \herd tool~\cite{herd-gh}.
The safety violation (left of ~\autoref{fig:litmuscode}) is due to a lack of synchronization in a release-acquire chain.
The intended synchronization between P0 releasing the lock and P2 acquiring the lock is broken if thread P1 interferes using a relaxed RMW operation like the CAS in \texttt{xchg\_tail}.
This is because the LKMM, unlike the C11 memory model, does not maintain release-acquire chains over relaxed RMW operations~\cite{LKMM-B-cumul}.
Changing the relaxed CAS to a release one (referred in Table~\ref{tab:veriftime} as \emph{``fix 2''}) solves this issue.
The await termination violation is due to a missing propagation which can make \texttt{arch\_mcs\_spin\_wait} loop forever.
While the \texttt{wmb()} in Line~\autoref{qspinlockcode:wmb} is enough to guarantee that the nodes are correctly initialized before changing the tail, in the LKMM, \texttt{wmb()} does not guarantee necessary propagation properties between different threads.
This problem can be solved by using \texttt{mb()} instead of \texttt{wmb()}.
The data race (right of~\autoref{fig:litmuscode}) occurs between the plain write to initialize
the MCS node (Line~\autoref{qspinlockcode:init} in~\autoref{fig:linux-cnacode-qspin}) and the write in \texttt{cna\_lock\_handoff} (Line~\autoref{qspinlockcode:handoff}) which is implemented using a \texttt{smp\_store\_release}.
There are two possible solutions to remove this race,
either initialize the node using \texttt{WRITE\_ONCE}, or use \texttt{mb()} instead of \texttt{wmb()} after the node initialization.
We opted for the latter since this single change (referred in Table~\ref{tab:veriftime} as \emph{``fix 1''})
solves both the await termination and data race problem.

With these two changes, \dartagnan finds no more issues in the core of \qspinlock.
However, it still reports data races in the critical section of the client code.
We deem those races to be spurious, because the fixed \qspinlock guarantees mutual exclusion and, hence, there cannot be a data race in the critical section according to any intuitive definition of data race.
As a result, we believe the data race definition of LKMM is broken.

\paragraph{Experimental Setup.}

The evaluation was executed on a GIGABYTE \textbf{R182-Z91-00} server~\cite{gigabyte-rack}
equipped with 2 \textbf{AMD EPYC 7352} 24-Core Processor~\cite{amd-epyc-processor} and 125 GiB of RAM.
Table~\ref{tab:veriftime} reports the verification time, the verification result and what kind of violations are found (if applicable).

\begin{figure}[t]
    \hspace{-8mm}
    \begin{subfigure}[t]{75mm}
        \input{figures/litmus-safety-code.tex}
    \end{subfigure}\hspace{5mm}
    \begin{subfigure}[t]{75mm}
        \input{figures/litmus-race-code.tex}
    \end{subfigure}
    \caption{Correctness violations in \qspinlock. Safety (left) and a data race (right).}
    \label{fig:litmuscode}
\end{figure}



\begin{table}
\centering
\caption{\dartagnan verification times for \qspinlock and CNA, with and without the applied fixes. We also report the verification results and what kind of violations are found.}
\label{tab:veriftime}
\begin{tabular}{clrcl}
\toprule
Memory model &               Lock algorithm & Verification time & Verified? & Violation type \\
\midrule
        LKMM &       \qspinlock, unmodified &              37 s &  \xmark{} &       Liveness \\
        LKMM &       \qspinlock, with fix 1 &             2 min &  \xmark{} &         Safety \\
        LKMM & \qspinlock, with fix 1 and 2 &             2 min &  \cmark{} &                \\
        \arm &       \qspinlock, unmodified &             3 min &  \cmark{} &                \\
      \power &       \qspinlock, unmodified &             3 min &  \cmark{} &                \\
        LKMM &              CNA, unmodified &            27 min &  \xmark{} &       Liveness \\
        LKMM &        CNA, with fix 1 and 2 &        15h 26 min &  \cmark{} &                \\
        \arm &              CNA, unmodified &        11h 42 min &  \cmark{} &                \\
      \power &              CNA, unmodified &        11h 23 min &  \cmark{} &                \\
\bottomrule
\end{tabular}
\end{table}

\section{Conclusion}

We have reported our efforts in verifying and optimizing the CNA lock on WMMs.
Using \vsync, we derived a safe and efficient set of barriers (with their associated memory order modes).
Using a client code with 3, 4, and 5 threads, we showed that the implementation is correct on WMMs with the maximally-relaxed sequence of barriers.

We also studied the CNA patch that is currently under review to be integrated into the Linux kernel.
Although we could not run our optimization on the patch, we could analyze it by applying few simplifications to the \qspinlock code.
We proved that the patch (v15) is correct under \arm and \power.
We described the violations found by \dartagnan under LKMM and explained how to fix them.

\bibliographystyle{plain}
\bibliography{references}
\end{document}